\documentclass[journal]{IEEEtran}

\usepackage{amsmath}              % for "\eqref" command
\usepackage{amsthm}
\usepackage{amssymb}
\usepackage{amsfonts}
\usepackage{graphicx}
\usepackage{latexsym,amssymb,epic,eepic}
\usepackage{xcolor}

\usepackage{hyperref}

\newcommand{\RR}{{\mathbb R}}%
\newcommand{\EE}{{\mathcal E}}%

\ifCLASSINFOpdf
\else
\fi
\DeclareMathOperator{\argmin}{argmin}
\DeclareMathOperator{\Vertex}{\mathcal{V}}
\DeclareMathOperator{\Triangle}{\Im}
\newcommand{\energy}{D}
\begin{document}
\title{Distortion measure of spectrograms for\\ classification of respiratory diseases}

\author{Jeremy~Levy,
        Alexander~Naitsat
        and~Yehoshua~Y.~Zeevi
\thanks{Jeremy Levy is with the Faculty of Electrical Engineering and the Faculty of Biomedical
Engineering, Technion---Israel Institute of Technology, Haifa 3200003 Israel}
\thanks{Alexander Naitsat and Yehoshua Zeevi are with the Faculty of Electrical
Engineering, Technion---Israel Institute of Technology, Haifa 3200003, Israel}}

\maketitle


%\begin{IEEEkeywords}
%IEEE, IEEEtran, journal, \LaTeX, paper, template.
%\end{IEEEkeywords}

% \fi
%\IEEEpeerreviewmaketitle

%%%%%%%%%%%%%%%%%%%%%%%%%%%%%%%%%%%%%%%%%%%%%%%%%%%%%%%%%%%%%%%%%%%%%%%%%%%%%%%%%%%%%%%%%%%
%\definecolor{changecolor}{rgb}{0,0, 0} 			   	   %	
\definecolor{changecolor}{rgb}{0,0, 0.8} 			   %
\definecolor{commentcolor}{rgb}{0.4,0, 0} 			   %
\definecolor{extracolor}{rgb}  {0.4,0.4, 0.4} 			   %

\newcommand{\marktext}[1]{\textcolor{changecolor}{#1}} % -->mark  changes made in the text
\newcommand{\markcomment}[1]{\textcolor{commentcolor}{#1}}  % 
\newcommand{\markextra}[1]{\textcolor{extracolor}{#1}}  % 
%%%%%%%%%%%%%%%%%%%%%%%%%%%%%%%%%%%%%%%%%%%%%%%%%%%%%%%%%%%%%%%%%%%%%%%%%%%%%%%%%%%%%%%%%%

\begin{abstract}
A new method for classification of respiratory diseases is presented.
The method is based on   a novel class of  features, extracted from pulmonary sounds, by parameterizing their spectrograms
that are represented as surfaces, and by utilizing geometrical distortions defined with reference to these surfaces.
This method yields a set of highly descriptive features for  analysis of pulmonary sound recordings. 
Furthermore, by combining these  features with  Mel-frequency cepstral coefficients, we introduce a powerful model
for automatic diagnosis  of  common respiratory pathologies.
Compared with baseline methods, our model achieves  superior results for  binary and multi-class classifications of common respiratory deceases. 
Our new approach to classification of one-dimensional signals is applicable to other signals
in the context of their representations in combined spaces or manifolds.
\end{abstract}

\section{Introduction}
\IEEEPARstart{M}{achine}
learning has begun to have impact on various facets of medical signals and image analysis
\cite{litjens2017survey,topalovic2019artificial}.
Computational means for lung sounds analysis have been the subject
of various studies using machine learning \cite{chamberlain_2016} and other related approaches \cite{Emmanouilidou_2012},
aiming at detecting  respiratory pathologies. 
%Since there is a limited number of well-trained physician \marktext{who specialize in osculation},
%the traditional  methods for detecting   respiratory pathologies are expensive and the diagnosis process is slow. 
%These methods  are often  subjected to human errors and the accuracy of  the diagnosis depends strongly on the physician experiences.

Nevertheless, the shortage in well-trained physicians who specialize in osculation pulmonary diseases is alarming.
This is especially the case in certain parts of the word like Africa, where the spread of obstructive lung diseases
is already beyond control \cite{pefura2016prevalence}.
Since advanced means of mass screening by medical imaging is not an option in such places,
let alone that expert radiologists are not available there, there is an urgent need for inexpensive,
yet reliable methods for automatic diagnostics of lung diseases by means of an electronic stethoscope.

%Furthermore, \marktext{spectral} attenuation due \marktext{both} the stethoscope \marktext{frequency response and auditory sensory limitations,} 
%may \marktext{compromise the quality of the final results.} 
% \marktext{Lastly, the fact that the} lung sounds are non-stationary signals, \marktext{render their analysis to be a challenging task}.

In view of the above outstanding medical problem, it is of utmost importance  to develop new algorithms for classification of
lung sounds and to advance a fully  automatic pipeline for diagnosis of common respiratory pathologies.

The physics of sounds and the natural characteristic of nonstationarity of lung sounds,
lend themselves to their analysis in the combined space of time-frequency, i.e. the spectrogram. 
We carry this approach one step further, by considering the spectrogram as a surface and by exploiting its geometric properties
in generating novel features that are powerful in classification of lung diseases by means of machine learning.

\section{Related work}
There exist several deep learning and model-based methods for automatic classification of lung pathologies 
based on their fingerprints that are hidden in the pulmonary sounds.
%from audio recordings.
For instance, the recent  method of \cite{imran2020ai4covid} implements a deep transfer learning-based multi-class classifier for
diagnosis of COVID--19, using cough recordings.

Chanbres et al. \cite{chambres2018automatic} employ the algorithm of the Essentia library \cite{bogdanov2013essentia} for extracting sound
features from cough recordings. 
This system was trained on   the dataset of the ICBHI 2017 challenge \cite{rocha2017alpha} and it uses a 
boosted-decisional-tree algorithm to classify sounds like crackles and wheezes.

Kandaswamy et al. \cite{Kandaswamy_2004} developed a method of analyzing lung sounds signals using wavelet transform,
combined with Artificial Neural Network (ANN) for the classification. They obtained an accuracy of 94.56\% on the
validation set and 91.33\% on the test set.
The dataset was composed of 126 recordings, categorized into inspiratory wheezes, fine crackles, stridor, squawk, and rhonchus,
apart from normal vesicular sounds.

A similar method with ANN layer was used  in \cite{amrulloh2015cough} for the analysis of cough sounds  recorded from pediatric patients.

Sankur et al.  \cite{Sankur_1994} proposed an auto-regressive (AR) model combined with KNN classifier for the task of
diagnosis of the lung sounds. They obtained Correct Classification (CC) of 93.75\% on the test set.
The dataset was composed of asthma, chronic bronchitis, emphysema, pneumonia, pleurisy and bronchiectasis pathologies.

\section{Methods}

\subsection{Database analysis}

The Respiratory Sound Database from (\url{https://www.kaggle.com/vbookshelf/respiratory-sound-database}) was used for the analysis.
A total of 918\ lung sounds recordings from 126\ patients were used. This database incorporates 7 different pathologies:
URTI, Asthma, COPD, LRTI, Bronchectasis, Pneumonia, Bronchiolitis, and healthy recordings. 

\begin{figure}[h!]
\centering
\includegraphics[width=1\columnwidth]{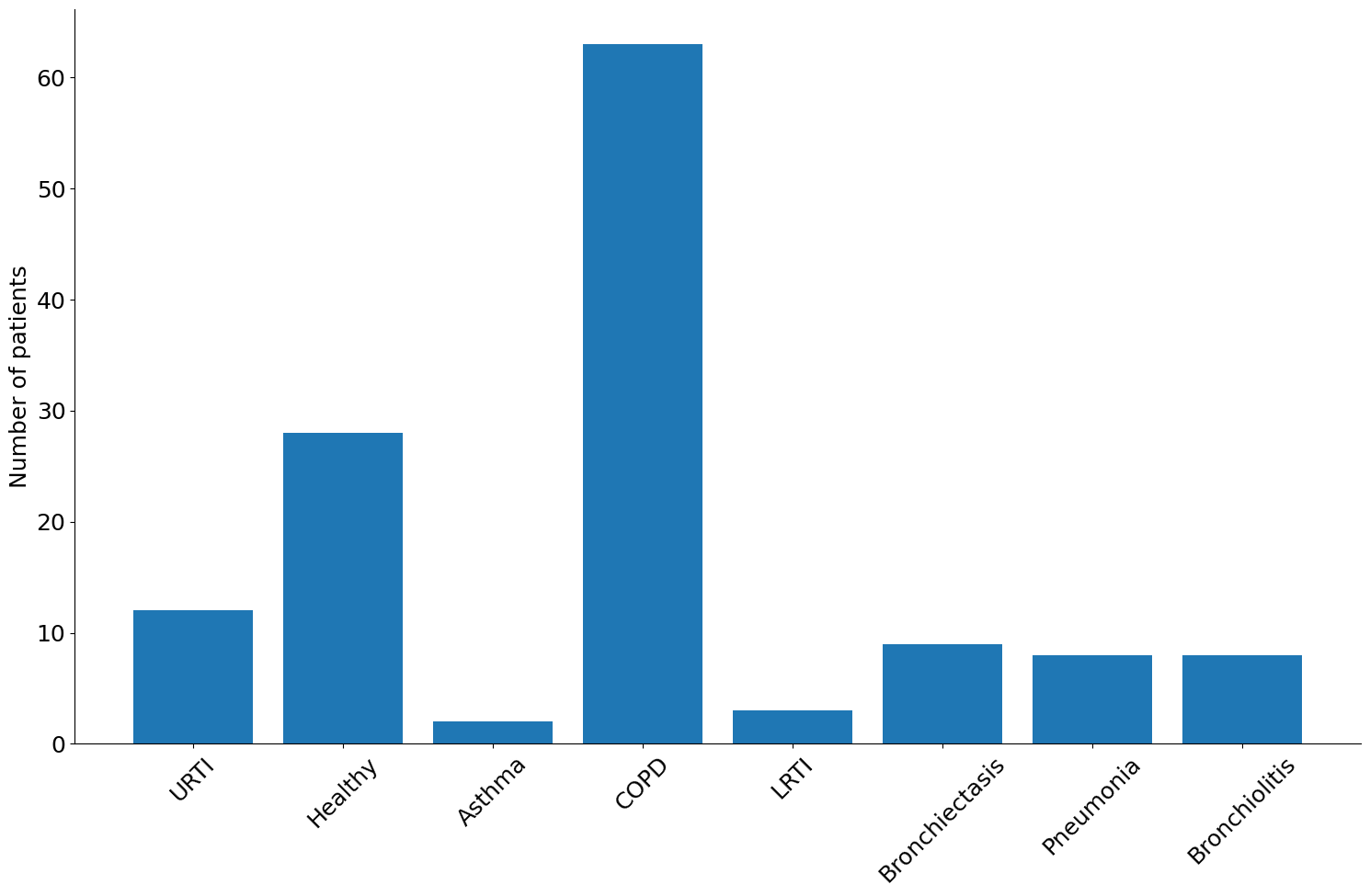}
\caption{Repartition of labels among the cases included in this database.
The database is imbalanced, about 50\% of the patients have COPD disease.}
\label{fig:database_analysis}
\end{figure}

The histogram depicted in Figure~\ref{fig:database_analysis} presents the repartition of the labels among the cases included in the database.
Due to the very poor occurrence of the  Asthma and LRTI pathologies, the corresponding recordings were excluded.

\subsection{Preprocessing}

One of the major problems that one must overcome in the process of analysis of lung sounds is $S/N$ level.
Sound generated by instruments  and other ambient activities affect significantly the quality of  the lung sounds signal. 
It is therefore crucial to improve the level of the $S/N$  without distorting the stethoscope's signal.

Our algorithm employs the Saviztky-Golay filter \cite{Savitzky} for denoising lung sounds. The purpose of this filter is to smooth the signal, and increase the SNR without altering the signal. This filter has been widely used in the field of time series analysis \cite{CHEN2004332}, especially for lung sound analysis \cite{HAIDER2018}.

The filter aims to fit a specific polynomial suitable for a signal frame, using least squares method. The central point of the window is replaced with that of the polynomial, producing a smoother signal.
Denote a polynomial of the degree $N$ by
\begin{equation*}
p(n) = \,\sum\limits_{k=1}^N\, a_k n^k \, ,
\end{equation*} then, the aim of  Saviztky-Golay filter is to minimize the following error:
\begin{equation*}
\EE_N = \,\sum\limits_{n=-M}^M\, \Bigl(p(n) - x[n])^2\Bigr) \,,
\end{equation*}
where $2M+1$ is the width of the window. %, $N$ is the order of the polynomial. 

\begin{figure}[h!]
\centering
\includegraphics[width=1\columnwidth]{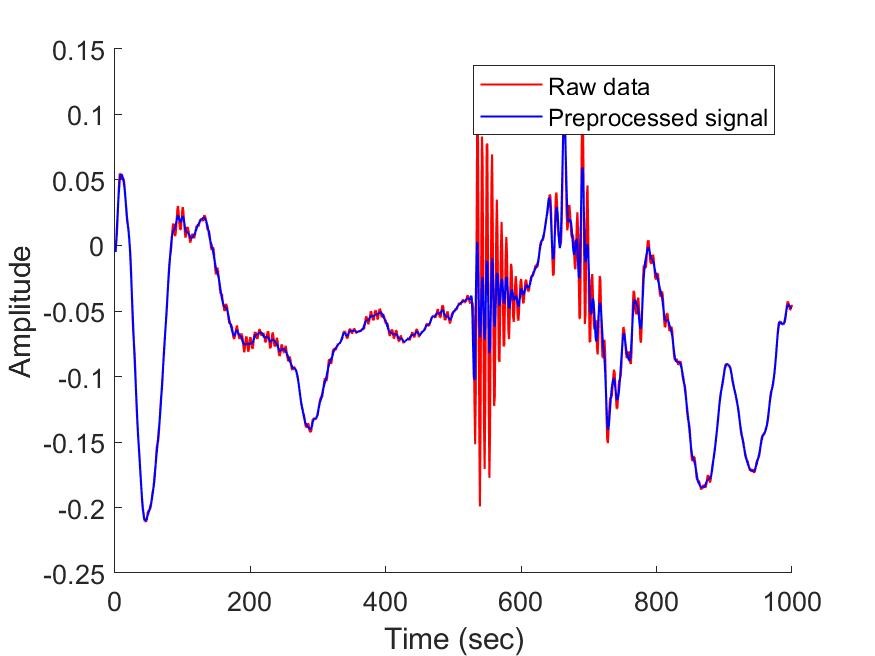}
\caption{Effect of the Saviztky-Golay filter on lung sounds recording.
Over the segment of the time period 530--580 sec., the signal is very noisy. %for about 50 seconds. 
The Saviztky-Golay filter  eliminates the  sharp changes.}
\label{fig:preprocessing}
\end{figure}

A large value of $M$ will produce a smoother signal, but may neglect some important variations in the signal.
A low value of $M$ may "over fit" the data. This is in a way the principle of uncertainty. 
Secondly, $N$, which specify the degree of the polynomial can produce a smooth signal for low value.
On the other hand, high value of $N$ may ``over fit'' the data. By experimenting with various combinations of these filter parameters,
we converged on the values of $N=3, M=11$ that yielded the best results.
Figure~\ref{fig:preprocessing} shows an example of a filtered signal, and the corresponding raw data. 

\subsection{Feature Engineering}
Our  algorithm  for the feature extraction can be decomposed into the two main parts: the signal processing and the geometry processing parts.

In the signal processing part, we preprocess  data and compute  spectrogram  images of the denoised audio signals.

In the geometry processing part, the original problem of signal processing is converted to another problem in which we analyze geometric representations of time-varying signals.
The first  goal of the geometry processing part is to represent  spectrogram  images  by discrete surfaces,
and the second goal is to reduce the problem of  signal processing to the problem of analyzing shapes of the obtained surfaces.
In this process, each sound signal is, first,  identified with its spectrogam surface, and then the measure of dissimilarity
between two signals is estimated as an effort that takes to deform one spectrogam surface onto the another.
	
By using various metrics for estimating this effort, we are capable of computing different geometrical quantities associated with spectrogram
surfaces.  We refer to these   quantities as to the \textit{geometrical distortions} or \textit{distortion measures}. 

In our method,  geometrical distortions 
are used  as descriptive features for analyzing  the original sound recordings. 
Based on these features, we train an automatic system for classification of respiratory diseases from sound recording.

Summarizing all the above, our algorithm for extracting features consists of the following steps:
\begin{enumerate}
\item
First, we use the FFT algorithm to  compute the spectrogram of each lung sounds recording.
Each recorded signal is represented  as a spectrogram surface $z(x,y)$, where $x$ is the time in seconds,
$y$ is the normalized frequency and $z$ is the corresponding  power-frequency (see Figure~\ref{fig:distortion_process}.  

\item
We next triangulate the  spectrogram  surfaces $z(x,y)$ using the  Delaunay  triangulation algorithm.
That is, we take $n$ samples $\Vertex=\big\{(x_1,y_1,z_1),\dots ,(x_n,y_n,z_n)\big\}$ of each spectrogram surface and compute
a set of triangles $\Triangle$ that connect vertices over $\Vertex$. 
The mesh of triangles obtained by that triangulation is denoted by  $(\Vertex,\Triangle)$.
It provides a discrete representation  of  the original sound signal  (see supplemental material for details). 

\item
Each mesh  $M=(\Vertex,\Triangle)$ is then mapped to the plane by a piecewise affine  map (simplicial map).
Here by  \textit{piecewise affine} or \textit{simplicial} map we refer to a map $f$ of a mesh $M$, defined in such a way that: \textbf{(i)} the restriction of $f$ to each triangle $t\in \Triangle$ is an affine transformation, denoted by $f_t$; \textbf{(ii)} components of  $f$ coincide on common edges  of  neighboring triangles.
[See examples of such mappings in Figure~\ref{fig:disotrtions_measures}.] 

\item 
Finally, we compute geometrical distortions induced by each simplicial map $f$ and use these distortions as descriptive features of
the original  signal (see  examples of color encoded distortions in Figure~\ref{fig:disotrtions_measures}).

\end{enumerate}

Apparently, the first two stages are aimed at computing a discrete geometric representation of audio signals,
whereas the goal of the last two stages is to compare these discrete representations.    
The  details of the above algorithm  stages are presented in the next sections.

\subsection{Sampling and Triangulation}
Assume that $S$ is a manifold surface, embedded in $\mathbb{R}^3$, and that $\Vertex$ is a finite set of points (vertices) sampled on $S$.
Then, a common way to discretize  $S$ is to divide that surface into a finite set of triangles $\Triangle$ such that: (\textbf{i}) vertices of the obtained triangles belong to $\Vertex$; $(\textbf{ii})$ for any pair of non-disjoint triangles $t_1,t_2\in \Triangle$ the intersection $t_1 \cap t_2$ is either a common edge of $t_1$ and $t_2$ or a common vertex of these triangles. We will refer to the pair $(\Vertex, \Triangle)$ as to the triangle mesh of a surface $S$.

In our case, each  input signal $I$ is represented by a spectrogram surface $S=S(I)$ that  can be written in following  parametric form:
\begin{equation}\label{eq:spectrogram_surfave}
S(I)=\big\{\big(x,y,z(x,y)\big) | x\in X, y\in Y \big\}\,,
\end{equation}
where $X$ and $Y$ are the time and frequency ranges of the audio signal $I$.
We divide $X$  and $Y$ into a number of uniformly  distributed  points $x_1,...,x_N$ and $y_1,...,y_N$, 
and  the vertex set $\Vertex$ of $S$  is defined by
\[ \Vertex =  \big\{(x_i,y_j, z(x_i,y_j))|\, i,j = 1,2,\dots,N \big\} \,. \]
We tested the above sampling method and found empirically  that it works  well for $N=150$.
However, in some scenarios, using adaptive sampling can potentially yield even better results.
See Section \ref{sec:discussion} for the discussion on  more advanced sampling schemes.

The triangle  set $\Triangle$ of $S$ is constructed by the standard algorithm for Delaunay triangulation \cite{edelsbrunner2001geometry}
that minimizes the minimum angle of all the angles of triangles in $\Triangle$. 
This triangulation  algorithm avoids generation of slim triangles which appearance may lead to numerical issues in the stage of
the feature extraction.

After representing data by triangular meshes, we proceed to the next step of analyzing geometric properties of these  meshes.    

\subsection{Shape descriptors} \label{sec:shape_descriptors}
Given two meshes of spectrogram surfaces,  we wish to define a metric suitable for of quantifying geometrical (de)similarities between these meshes. 
In computer vision, such metrics are often  referred to as \textit{shape descriptors}.
There exist many approaches  to computing shape descriptors for a collection of 3D objects.
These approaches can be divided qualitatively into the following categories:
\begin{enumerate}
	
\item \textbf{Spectral methods}.
In spectral approaches, shape descriptors are derived from  discrete representations of the  Laplace-Beltrami operator, defined on surfaces \cite{reuter2006laplace}. 
Cotangent weights   are most commonly used for approximating Laplace-Beltrami operators over meshes. 
By using cotangent weights,  the Laplace operator action  on a mesh $M$ can be represented by a sparse Laplacian matrix  $L=L(M)$.
In such a case, the spectral descriptors of $M=(\Vertex,\Triangle)$ are often defined as $n$-largest eigenvalues of $L$,
for a constant number $n < |\Vertex|$ \cite{rustamov2007laplace}.

\item \textbf{Metric  methods}. 
These methods  represent each mesh $M$ by a matrix $G$ of pairwise  distances between vertices of $M$. Usually, these are the Euclidean or geodesic distances. 
A desimilarity measure between two meshes $M_1$ and $M_2$ is defined in the metric approaches  as a function of the distance matrices $G_1$ and $G_2$ of these two meshes.  For example, metric descriptors of triangulated surfaces  can be obtained by solving the problem of the  General Multi-Dimensional Scaling (GMDS) \cite{bronstein2006generalized},  or by solving  other related problems that involve computations of geodesic distances \cite{hamza2003geodesic,elad2003bending}.
	
\item \textbf{Deformation-based  methods}. 
In deformation-based methods,  a distance between two shapes $S_1$ and $S_2$ is estimated by computing an optimal deformation
$f_{12}$ of $S_1$ onto  $S_2$ and by measuring changes in various  geometric features induced by $f_{12}$.
There exist many criteria for definition of map's optimality. Most of these criteria are targeted at preserving the map injectivity
and avoiding visual distortions, as much as possible.
	
Note that for a large collection $\{S_1,...,S_m\}$ of shapes  it may be very expensive to compute optimal  deformations $f_{ij}$,
for each $1 \leq i<j\leq m$.  
Therefore, instead of matching all the pairs of shapes, a more practical approach would be  to compute an optimal mapping $f_i$ of each shape $S_i$ into a simple target domain.
Depending on the dimensionality of an object and its topology, a simple target domain could be the  plane, a sphere \cite{rustamov2013map},
the unit circle \cite{boyer2011algorithms},  a cube \cite{sandhu2012volumetric}, or a ball \cite{NAITSAT201837}.
\end{enumerate} 

Our model employs  deformation-based descriptors for measuring similarities between triangle meshes.
Note that all the meshes that constitute a peak surface of spectrograms have the topology of a planar disc. 
Therefore, a natural candidate for the optimal deformation of a such  mesh $M$ is a length-minimizing mapping of $M$ into the plane.
We refer to this mapping process as to the \textit{surface flattening}, for short. 
In our model, surface flattening algorithms are used  for computing deformation-based  descriptors  of spectrographic shapes. 

If $f$ is a flattening of a mesh $M$, then we select the shape descriptors of $M$ to be the geometrical distortions that measure
how Euclidean lengths are deformed under $f$. 
In such a way, each mesh $M$ can be associated with its  signature vector  $(E_1,...E_2)$, where numbers $E_i$ are various
estimates of the metric deviations induced by flattening $M$ into the plane.

In the the sequel, we address in the  detail the  surface flattening and the distortion estimation processes.

\color{black}

\subsection{Surface flattening}
The problem of flattening  triangular meshes into the plane, also referred to as  the parametrization problem, constitutes one of the central issues in geometry processing. Consequently, there exist many algorithms for flattening triangulated surfaces\footnote{See \cite{hormann2007mesh,Naitsat2020_preprint} for the survey of parametrization algorithms and for other related methods used in geometry processing.}.
These algorithms are  aimed at computing a locally-injective parametrization that minimizes distortions of fundamental geometric quantities,  such as angles and lengths.
Surface parametrization tasks can be  reduced to the following optimization problem:
\begin{equation} \label{eq:parameterization_problem}
\begin{aligned}
f^*=~ &\underset{f}{\argmin}~ E(f); \\
\text{s.t.~~~~}&\det df_t > 0,  t\in \Im\,,
\end{aligned}
\end{equation}
where $f^*$ is a piecewise affine mapping of a mesh $(\Vertex,\Triangle)$ that minimizes the chosen distortion criteria $E$
under the following constraints:  for each mesh triangle $t$, the component of $f^*$ on $t$ is an orientation preserving  map. 
These constraints are expressed by the determinant signs of Jacobian matrices $df_t$, $t\in \Triangle$.   Negative determinants
of the Jacobians yield inverted triangles in the image of $f$. Therefore, satisfying the orientation constraints is the necessary condition
for inducing  one-to-one parametrization of surface meshes.

We adopt the recently-proposed Adaptive Block Coordinate Descent (ABCD) algorithm \cite{ABCD}, combined with the Tutte
embedding method \cite{Floater_2002}, to solve the optimization problem \eqref{eq:parameterization_problem} and thereby the
parametrization problem.
In particular, we initialize the parametrization problem \eqref{eq:parameterization_problem} by mapping triangular meshes onto a circle
via the method of  \cite{Floater_2002}. 
We then employ  the ABCD algorithm to induce  locally injective parametrization characterized by minimal length distortions.
 
Note that, since  \eqref{eq:parameterization_problem} is a non-convex problem, 
solving it with different initial maps  may lead to distinct local minima.
Therefore, choosing an appropriate initialization method  is crucial for adequate  approximation  of the global minimizer $f^*$.

\begin{figure}[h!]
\centering
\includegraphics[width=0.5\textwidth]{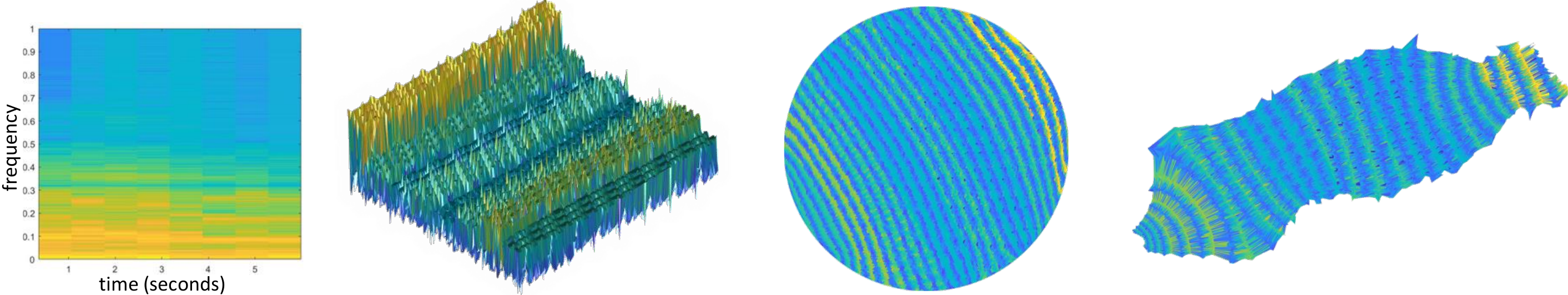}
\caption{Visualization of the process of measuring distortion measures for a given lung sound segment. 
The figure shows from  left to right: a sound  spectrogram, a triangular mesh $(\Vertex,\Triangle)$ representation of the spectrogram surface,
an initial mapping of  $(\Vertex,\Triangle)$ onto the plane and the final flattening of the mesh, computed by means of the ABCD algorithm.}
\label{fig:distortion_process}
\end{figure}

We tested a number of different initialization schemes, and found that using a convex combination mapping of meshes \cite{Floater_2002}
onto a planar disc, yields the best results. 
Note that the  algorithm of \cite{Floater_2002} is actually  a variant of the classical Tutte embedding algorithm  that 
is widely used in shape processing applications. 
This method guarantees  a bijective mapping onto convex planar domains and it has a low  computational cost.
Figure~\ref{fig:distortion_process} demonstrates this  initialization scheme  and the related  process of distortion minimization.

We proceed to discuss the process of feature extraction. It includes the \textit{local sub-step} of extracting features of individual triangles and the \textit{global sub-step} in which local features are summed over large subsets of mesh triangles.

\subsection{Measuring local distortions} \label{sec:local_distortions}
If $M=(\Vertex,\Triangle)$ is a triangle mesh and $f$ is a simplicial mapping of $M$, then a \textit{local distortion} induced by $f$, on a triangle $t$, is defined to be a function $E(\sigma_1,\sigma_2) $ of the singular values $\sigma_1(df_t)$ and  $\sigma_2(df_t)$ of the Jacobian $df_t$.

The Jacobian singular values uniquely define the shape of a  triangulated surface, up to rotation and sliding of mesh triangles. 
Generally speaking, local distortions   estimates  how extensively is the shape of $t$ distorted under $f$.

These measures are instrumental in many applications in computer vision, including shape classification and shape analysis
\cite{NAITSAT201837,Chen_2008}. In our algorithm, geometric distortions are used  as measures of dissimilarity of triangulated surfaces\footnote{In the context of the lung sound representation on the surface of the local spectrum, the geometric  distortions assess how much the local spectrum of the sound is affected by the simplicial mapping $f$.}.

Note that for a dense triangulation, feeding  singular values  $ \big\{ \sigma_i(df_t)| t \in \Triangle,\, i=1,2\big\}$  to a deep learning model preserves all the information contained in the pixels of the spectrogram.     	

Our algorithm employs several distortion measures. These distortions belong to the following major classes of geometric measures:

\noindent
{\bf Isometric distortions.} These measures estimate distortions of the Euclidean length.
We use the following isometric distortions:
\begin{itemize}
\item
ARAP (As-Rigid-As-Possible) energy  \cite{Sorkine_2007}
\begin{equation*}
E_{\mathrm{ARAP}} (\sigma_1, \sigma_2 ) = (\sigma_1^2 -1)^2 + (\sigma_2^2 -1)^2\,;
\end{equation*}
\item
Symmetric Dirichlet energy \cite{Smith_2015}
\begin{equation}
E_{\mathrm{SD}} (\sigma_1 , \sigma_2 ) = \displaystyle\frac{1}{4} \,
 (\sigma_1^2 + \sigma_1^{-2} + \sigma_2^2 + \sigma_2^{-2})\,;
 \label{eq:E_SD}
\end{equation}
\item
Quasi-isometric $ (qi) $ dilatation \cite{Naitsat_2016},\cite{Sorkine_2002}
\begin{equation*}
E_{\mathrm{QI}} (\sigma_1 , \sigma_2 ) = \max\limits \, \{ \sigma_1 , \sigma_2^{-1}\}\,;
\end{equation*}
\end{itemize}

\noindent
{\bf Conformal distortions.} 
These distortions estimate how far $f$ is from being an angle-preserving mapping.
Our algorithm uses the following estimates of conformal distortions: 
\begin{itemize}
\item
Quasi-conformal \textit{(qc)} dilatation \cite{Naitsat_2014}
\begin{equation}
E_{\mathrm{QC}} (\sigma_1 , \sigma_2 ) = \max\limits \, \left\{\displaystyle\frac{\sigma_1}{\sigma_2},
\frac{\sigma_2}{\sigma_1}\right\}\,;
\end{equation}
\item
MIPS energy \cite{MIPS_1999},\cite{Fu_2015} 
\begin{equation}
E_{\mathrm{MIPS}} (\sigma_1 , \sigma_2 ) = \displaystyle\frac{\sigma_1}{\sigma_2} +\displaystyle\frac{\sigma_2}{\sigma_1}
= \displaystyle\frac{\sigma_1^2 + \sigma_2^2}{\sigma_1 \sigma_2}\,;
\label{eq:E_MIPS}
\end{equation}
MIPS (Most Isometric Parametrizations) is a quadratic function, widely used for optimizing conformal distortions over
triangular domains \cite{Fu_2015}.
\end{itemize}

\noindent
{\bf Area distortions.} 
These distortions estimate dilatation and compression of triangle areas induced by $f$.
We use the following measure of the area distortion:
\begin{itemize}
\item
Unsigned area distortion \cite{Degener_2003}
\begin{equation}
E_{\mathit{AD}}(\sigma_1,\sigma_2)  = \max\limits \, \Bigl\{ |\sigma_1  \sigma_2| , | \, \sigma_1  \sigma_2 |^{-1}\Bigr\}\,;
\label{eq:E_AD}
\end{equation}
\end{itemize}

\noindent
{\bf Scale distortions.} These distortions assess the degree to which mesh triangles are scaled by $f$.
Scale distortions are closely related to discrete harmonic mappings \cite{Ezuz_2019} and to stretch-minimization mappings \cite{Sander_2002}.
We use the following scale distortions:
\begin{itemize}
\item
\textit{Dirichlet} energy \cite{MIPS_1999}
\begin{equation}
E_{\text{Dirichlet}} \, (\sigma_1 , \sigma_2) = \frac{1}{2}\, \Bigl(\sigma_1^2 + \sigma_2^2\Bigr)\,;
\end{equation}

\item
Conformal factor \cite{Chen_2008} 
\begin{equation}
E_{\text{CF}}(\sigma_1,\sigma_2) = \, \displaystyle\frac{\sigma_1 + \sigma_2}{2}\,.
\end{equation}

Note that conformal factors are closely related to conformal distortions such as quasi-conformal dilatation and MIPS energy.
Indeed, according to the uniformization theorem \cite{abikoff1981uniformization} any disc-topology surface $S$ can be mapped into the
plane by a conformal map $f_S$. The map $f_S$ can be described by its conformal factors,
up to a composition of $f_S$ with a rigid transformation. 
For this reason, the conformal factor has been used by \cite{Chen_2008} as a geometric signature for a collection of 3D surfaces.
\end{itemize}

All those distortion measures are rotation invariant, since they are functions of signed singular values of the Jacobian.
This work aims to show that the dimensionality of the data can be considerably reduced by employing 
weighted sums of local distortions over different subsets of $\Triangle$. 
The obtained quantities will be referred  to as a \textit{global distortions}.

\subsection{Measuring global distortions}
Let $f$ be  a simplicial map of the mesh $(\Vertex,\Triangle)$, $E$ be a local distortion, $\Triangle_0$ be a subset of $\Triangle$. The global distortion of $f$, computed with respect to $E$ over $\Triangle_0$, is then defined as follows:
\begin{equation}
	\energy_{\Triangle_0}(f, E) = \displaystyle\frac{\sum\limits_{t\in \Triangle_0} \,  E\big(\sigma_1(df_t),\sigma_2(df_t) \big)  \text{area}(t)}{\sum\limits_{t \in \Triangle_0} \, \text{area}(t)}\,,
	\label{eq:global_distortion}
\end{equation}
where   $df_t$  is the Jacobian of $f$ on $t$,  $\sigma_1(df_t)$ and  $\sigma_2(df_t)$ are the Jacobian singular values 
and $\text{area}(t)$ denotes the area of a triangle  $t$.

\begin{figure}[h!]
\centering
\includegraphics[width=0.5\textwidth]{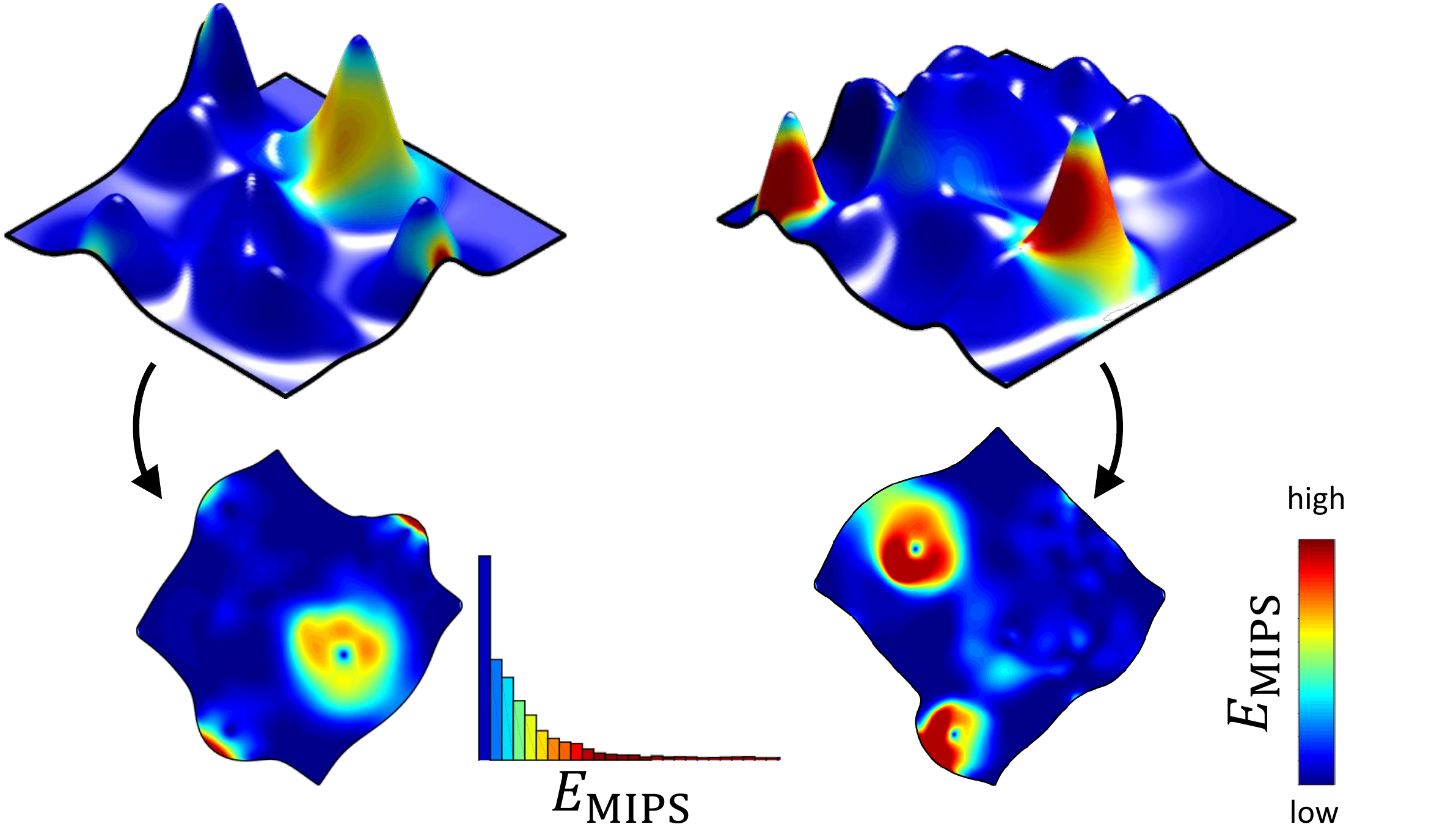}
\caption{Flattening surfaces  and measuring the resultant  geometrical distortions, attained as shape descriptors.
The process is visualized for the MIPS distortion  energy, defined by \eqref{eq:E_MIPS}. 
We first use ABCD algorithm  \cite{ABCD}, initialized with  Tutte embedding, to map  the triangulated surfaces into the plane.
We then compute  Jacobian  singular values $\sigma_1(t)$ and $\sigma_2(t)$  over each mesh triangle $t$.
Finally, for each distortion measure $E(\sigma_1,\sigma_2)$ and mapping $f$ we compute the two quantities $E_1(f)$ and
$E_2(f)$, defined according to \eqref{eq:global_distortion} and \eqref{eq:E_12(f)}.
} \label{fig:disotrtions_measures}
\end{figure}
\begin{figure}[h!]
\centering
\includegraphics[width=1\columnwidth]{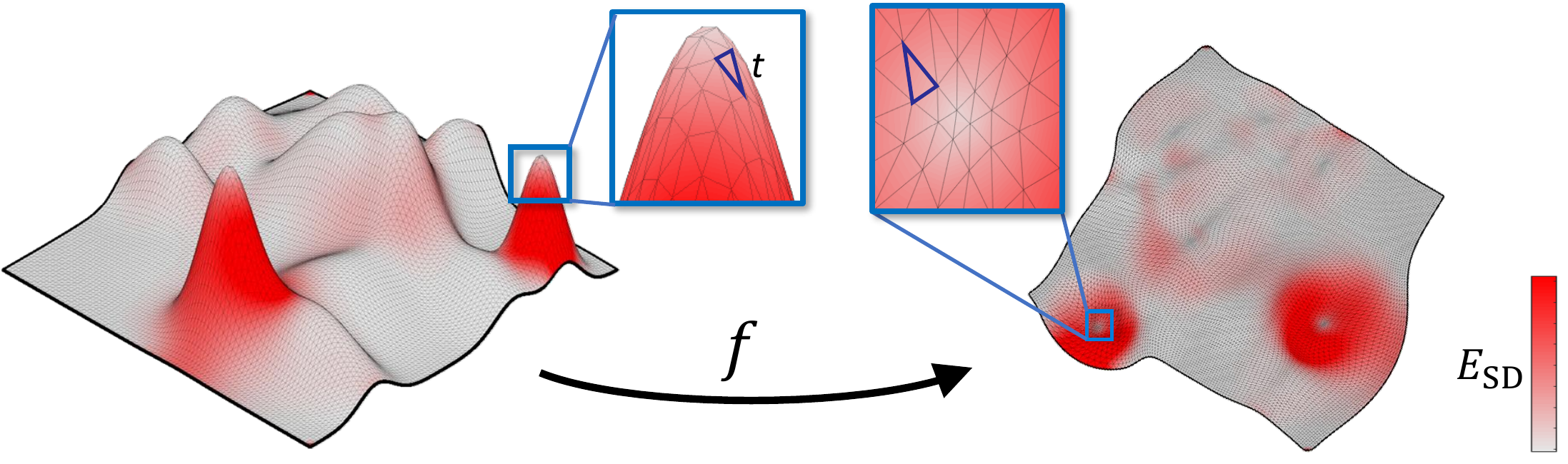}
\caption{Visualization of a  simplicial map $f$  with color encoded local distortions $E_{\text{SD}}$ , defined by \eqref{eq:E_SD}. 
At the top,  we  show how an individual  triangle $t$  is mapped under an affine  component $f_t$ of the map $f$. 
}
\label{fig:simplicial_map}
\end{figure}

In many cases, values of local distortions are distributed non-uniformly over mesh triangles. As demonstrated by Figures \ref{fig:disotrtions_measures} and \ref{fig:simplicial_map},
a small number of highly distorted triangles may have more impact on the global distortion $\energy_{\Triangle}(f, E)$ than the rest of the mesh triangles.
Therefore, in order to extract more information from each distortion measure,  one  can divide the triangle set $\Triangle$ into  a number of disjoint subsets. 
We employ this approach to extract more features for each distortion measure $E(\sigma_1,\sigma_2)$ and to compensate for the
adverse effects of a non-uniform distribution of distortions. 
In particular, we divide triangles into the  two subsets  according to \textit{triangle frequency}.

The frequency of a triangle $t$ is defined as the frequency of the center of gravity of $t$. The median over the
frequencies of all the triangles of the surface is computed. Then, the global distortion $\energy_{\Triangle_1}(f, E) $ is computed for all the
triangles with a frequency lower than the median ($t\in \Triangle_1$), and a second global distortion  $\energy_{\Triangle_2}(f, E) $ is computed  for the triangles with a frequency greater that the median ($t\in \Triangle_2$).
In such a way, each distortion measure $E$  contributes the following two features: $D_{\Triangle_1}(f,E)$ and $D_{\Triangle_2}(f,E)$. 
	We will denote these  features by  $E_1(f)$ and $E_2(f)$, for short.  
That is,
\begin{equation} \label{eq:E_12(f)}
	E_i(f) =  D_{\Triangle_i}(f,E),~ i=1,2\,,
\end{equation}
where $D_{\Triangle_i}$ is defined according to \eqref{eq:global_distortion}.

To summarize, we measure  global distortions  over the two subsets of triangles and use the obtained quantities as  
 a shape descriptors of spectogram surfaces.
 This approach  has the following advantages over the previously available distortion-based models for shape analysis 
 \cite{Chen_2008,NAITSAT201837}:

\begin{enumerate}
\item A wider set of distortion measures is used.
\item 	The overall number of features is further increased by  dividing distortions into the low and high frequencies.
\item The method operates on triangular meshes instead of tetrahedral meshes. Compared with the volumetric method of \cite{NAITSAT201837},  extracting features in our algorithm has a lower computation cost\footnote{We use  triangular meshes  because our data is represented by disc-topology surfaces. However, it is possible to represent this data by tetrahedral meshes and to estimate volumetric distortions of these meshes. See the Discussion section for more details.}.
\end{enumerate}

%\subsection{Statistical analysis}

%To evaluate whether an individual feature is discriminative for the classification, a Kruskal-Wallis test has been used, with
%post-hoc analysis. The test is non-parametric and evaluates the null hypothesis that the two distributions have equal medians.
%If $ p<$5\%, then the test rejects the null hypothesis of equal medians at the default 5\% significance level.
%The test has been used for each distortion measure.

\begin{figure}[h!t!b!]
\centering
\includegraphics[width=1\columnwidth]{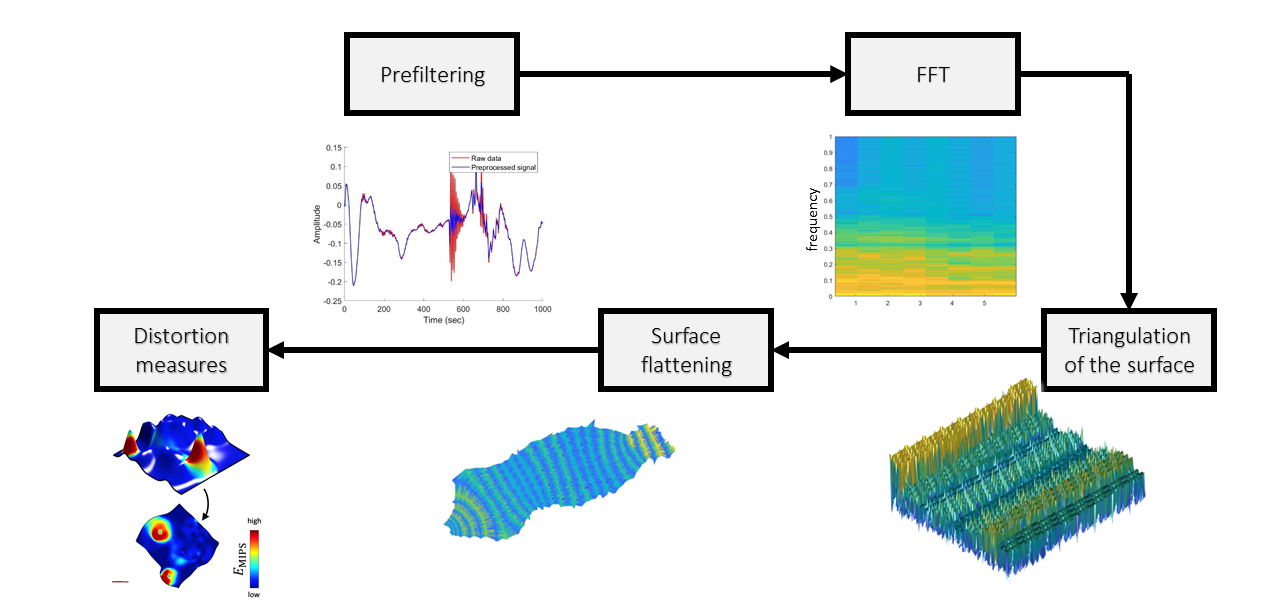}
\caption{High-level architecture of the proposed model. The model is composed of three main steps: signal processing, geometric learning and classification by means of machine learning.}
\label{fig:model_overview}
\end{figure}

A high-level overview of the proposed model is summarized in Figure~\ref{fig:model_overview}.

\subsection{Baseline}

A baseline (i.e. a reference mode) has to be created for comparison the model created with. A different approach has been selected for this purpose, based on a set of features has been  handcrafted. Twelve Mel Frequency Cepstrum Coefficients (MFCCs) were extracted from the audio files. MFCC is the most widely used feature extraction method in automatic speech recognition \cite{OSHAUGHNESSY20082965}. In the feature extraction phase, 6 statistical parameters have been extracted from each of the 12~MFCC coefficients as follows: mean, standard deviation, min, max, mean of the absolute difference, standard deviation of the absolute difference. Altogether,  72~features. 

The reference model has been applied to  all  classifiers, with the same training cost in the case of the proposed model. The models that we use for comparison have been trained and tested on the same train/test subdivision of the data.
The proposed model was first applied with each distortion measure separately, to assess the efficiency of each of them.
Combination of the distortion measures were then tested.

Finally, the MFCC-based model has been combined with the proposed model (based on distortion measures).
For each recording, 88~features have been computed: the 16 features based on distortions measures (2 for each distortion), and the
72~features based on MFCC coefficients. As the number of features increased significantly, a feature selection step has been applied,
based on the ranking of features, determined by  implementation in the Random Forest classifier. Altogether, 45~features have been selected.

\section{Implementation}

\subsection{Training}

A total of two classification tasks have been conducted: a multi-class classification, with the 5 pathologies and the healthy recordings, and a binary classification, for each of the 5~pathologies against the class of healthy recordings.

The dataset was subdivided into 80\% training set and 20\% test set.
For each task, several classifiers were experimented: Logistic regression (LR), support vector machine (SVM), Random forest (RF), $K$ nearest neighbors (KNN), and AdaBoost (AB).
For all of these models we used the 16 engineered features. For each model, hyper-parameters such as the number of estimators or number of neighbors were optimized using 5--fold cross validation. A large random grid of hyper-parameters was searched for (see Supplements). In the case of the multi-class classification, the performance measure used for optimization was the accuracy, whereas for the binary tasks the area under the receiver operating characteristics curve (AUROC) was used. A weight has been assigned to each class, inversely proportional to the class frequencies in the training set.

For each iteration of the cross-fold, training examples were divided into training and validation set by stratifying among patients, which means several recordings from the same patient are always in the same set.

\subsection{Evaluation}

All the models were trained on the same test set. That is,  for all the models, the database was split into the same training and test subsets.

The following metrics were used for the performance evaluation:

$$ Accuracy = \frac{TP + TN}{TP + TN + FP + FN}\,,$$
$$ Recall = \frac{TP}{TP + FN}\,,$$
$$ Jaccard = \frac{TP + TN}{2*(P+N) - (TP + TN)}\,,$$
where $TP, TN, FP, FN$ are respectively the the True Positives, True Negatives, False Positives and False Negatives. $P$ is the number of positives samples, $N$ is the number of negatives samples.
\newline The  area under the ROC curve, $AUROC$, is computed.

\section{Results}
\subsection{MFCC baseline model}
The results of the MFCC baseline model are summarized in Table~\ref{table:MFCC},

\begin{table*}[h!]
\renewcommand{\arraystretch}{1.1}
\centering
\caption{Results of the binary classification utilizing the MFCC-based model.}
\label{table:MFCC}
\begin{tabular}{lccccc}
\hline
& URTI & COPD & Bronchiectasis & Pneumonia & Bronchiolitis\\\hline 
\textit{$AUROC$} & 0.82 & 0.98 & 0.95 & 0.90 & 0.81\\ 
\textit{Accuracy} & 0.83 & 0.99 & 0.96 & 0.91 & 0.70\\ 
\textit{Recall} & 0.83 & 0.99 & 0.96 & 0.91 & 0.70\\ 
\hline
\end{tabular}
\end{table*}

\subsection{Classification}

The results obtained by each classifier are summarized in Table~\ref{table:five_classifiers} and the results of the binary classification appear in Table~\ref{table:binary_classifier}.
For clarity, we report only the best results of the binary classifications in Table~\ref{table:binary_classifier}.

\renewcommand{\arraystretch}{1.1}
\begin{table}[h!]
\centering
\caption{Results obtained the 5 classifiers on the test set. The best score for each metric is underlined.}
\label{table:five_classifiers}
\begin{tabular}{lccc}
\hline
Classifier& \textit{Accuracy} & \textit{Jaccard score} & \textit{Recall}\\\hline 
KNN & 0.86 & 0.76 & 0.86\\
LR & \underline{0.88} & \underline{0.80} & \underline{0.89}\\
AB & \underline{0.88} & \underline{0.80} & \underline{0.89}\\
RF & \underline{0.88} & 0.79 & 0.88\\
SVM & \underline{0.88} & 0.78 & 0.88\\
\hline
\end{tabular}
\end{table}

\begin{table*}[h!]
\renewcommand{\arraystretch}{1.1}
\centering
\caption{Score for binary classifiers, for the 5~different pathologies.}
\label{table:binary_classifier}
\begin{tabular}{lccccc}
\hline
& URTI & COPD & Bronchiectasis & Pneumonia & Bronchiolitis\\\hline
\textit{$AUROC$} & 0.89 & 0.99 & 1.00 & 0.87 & 1.00\\ 
\textit{Accuracy} & 0.90 & 0.97 & 1.00 & 0.87 & 1.00\\ 
\textit{Recall} & 0.90 & 0.97 & 1.00 & 0.87 & 1.00\\ 
\hline
\end{tabular}
\end{table*}
\begin{table*}[h!]
\renewcommand{\arraystretch}{1.1}
\centering
\caption{Results of the combined model, for classifying 5 pathologies.}
\label{table5}
\begin{tabular}{lccccc}
\hline
 & URTI & COPD & Bronchiectasis & Pneumonia & Bronchiolitis\\\hline 
\textit{$AUROC$} & 0.93 & 1.00 & 1.00 & 1.00 & 1.00\\ 
\textit{Accuracy} & 0.92 & 1.00 & 1.00 & 0.90 & 1.00\\ 
\textit{Recall} & 0.92 & 1.00 & 1.00 & 0.90 & 1.00\\ 
\hline
\end{tabular}
\end{table*}

\begin{figure}[h!]
\centering
\includegraphics[width=0.6\columnwidth]{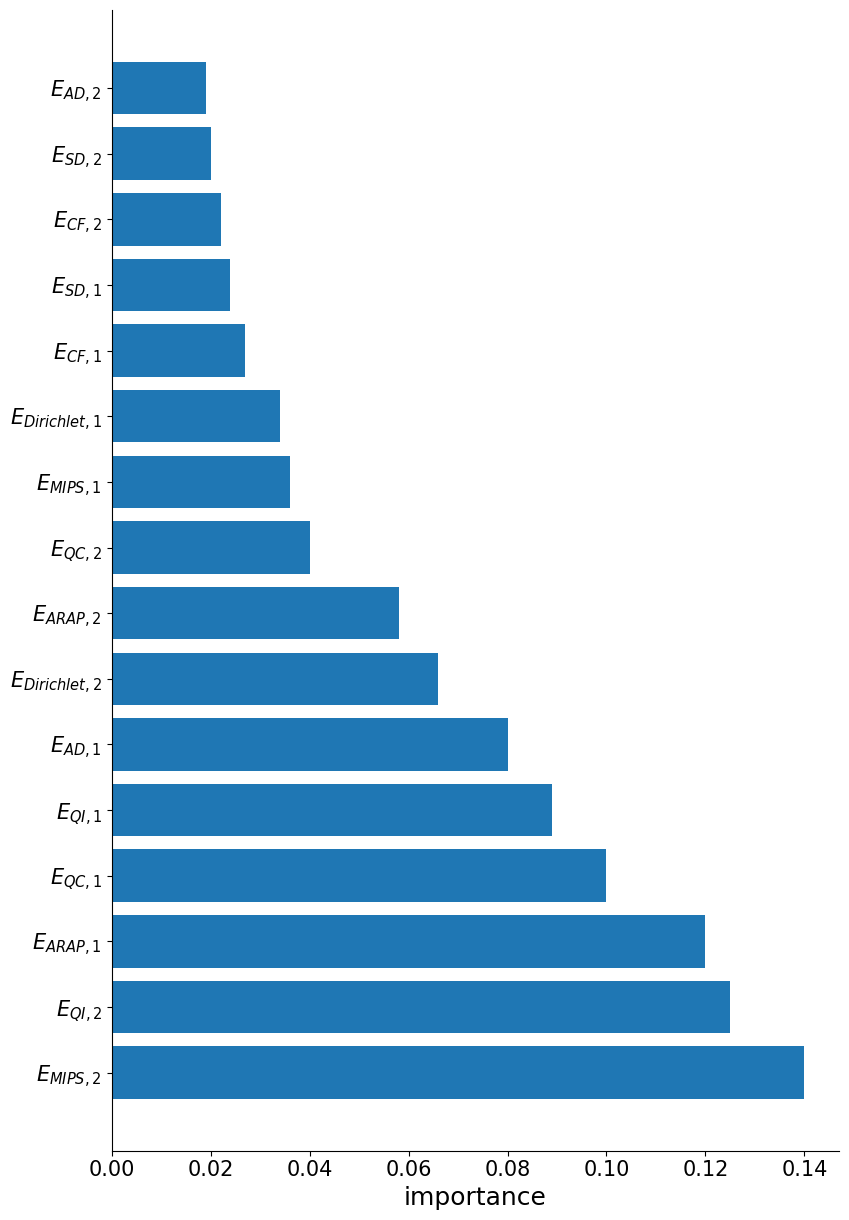}
\caption{Feature importance, determined by  using a Random Forest classifier. The ranking of the features allows us to understand what are the most useful features in the classification, and yields further insight into  the model.}
\label{fig:feature_importance}
\end{figure}

Ranking of the features according to their importance, as determined by Random Forest classifier, is depicted in Figure~\ref{fig:feature_importance}.

The proposed model obtains a better $AUROC$ than the baseline models for  almost all the binary tasks. For differentiation  of pneumonia pathology  from the rest of diseases, the proposed model yields a lower $AUROC$ value than the baseline model (0.87 vs 0.90). %Indeed, the statistical analysis showed that the features are less discriminative for this pathology.

\subsection{Combination of models}
\begin{figure*}[h!]
\centering
\includegraphics[width=1.7\columnwidth]{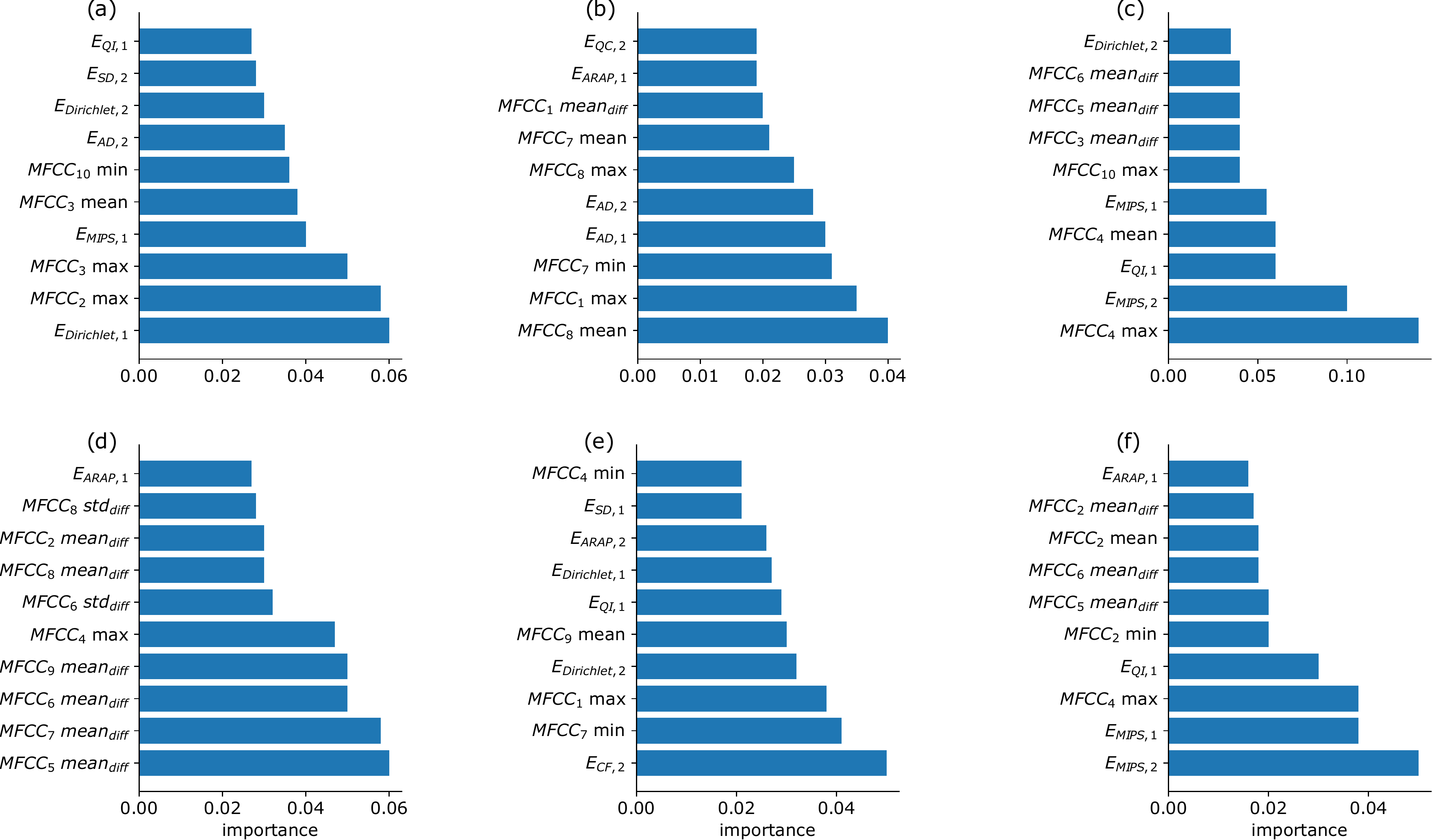}
\caption{Ranking of the 10 best features of the combined model, determined by means of the performance of the Random Forest classifier. On panel (a), the ranking is presented for the binary task of identifying  Bronchiectasis. On panel (b), for the Bronchiolitis. On panel (c), for COPD. On panel (d), for Pneumonia. On panel (e), for URTI. Finally on panel (f), the ranking is presented for the multi class task of classifying the 5 pathologies.}
\label{fig:ranking_all}
\end{figure*}

After comparing the proposed models with the MFCC-based model, a combination of the models has been implemented as follows:

For each recording the feature vectors from the 2 models have been concatenated, leading to a total of 88 features.
Figure~\ref{fig:ranking_all} presents the ranking of the features, for each of the 5~binary tasks and the multiclass task of identifying the 5~pathologies. Although there are 16~distortion measures features and 72~MFCC features, for most of the pathologies the occurrence of the distortion measure features is relatively high. In particular, there are six distortion  measures out of ten most highly-ranked features for the Bronchiectasis and URTI pathologies. Likewise, distortion measures appear among the four most highly-ranked features used in classification of the Bronchiolitis and COPD diseases. Indeed, for this pathology the MFCC-based model outperformed the  proposed model. However, if our task is to identify a Pneumonia lung sound, then only a single distortion measure appears in the feature ranking list. Indeed, for this pathology the MFCC based model outperformed the proposed hybrid model. 

Secondly, Table~\ref{table5} presents the results of the combined model for the  binary tasks.
This hybrid  model receives a $AUROC$ of 1.00 for Bronchiectasis, Bronchiolitis and COPD diseases.
To summarize, the combination of both approaches with the feature selection step improved the results.

\begin{figure}[h!]
\centering
\includegraphics[width=.7\columnwidth]{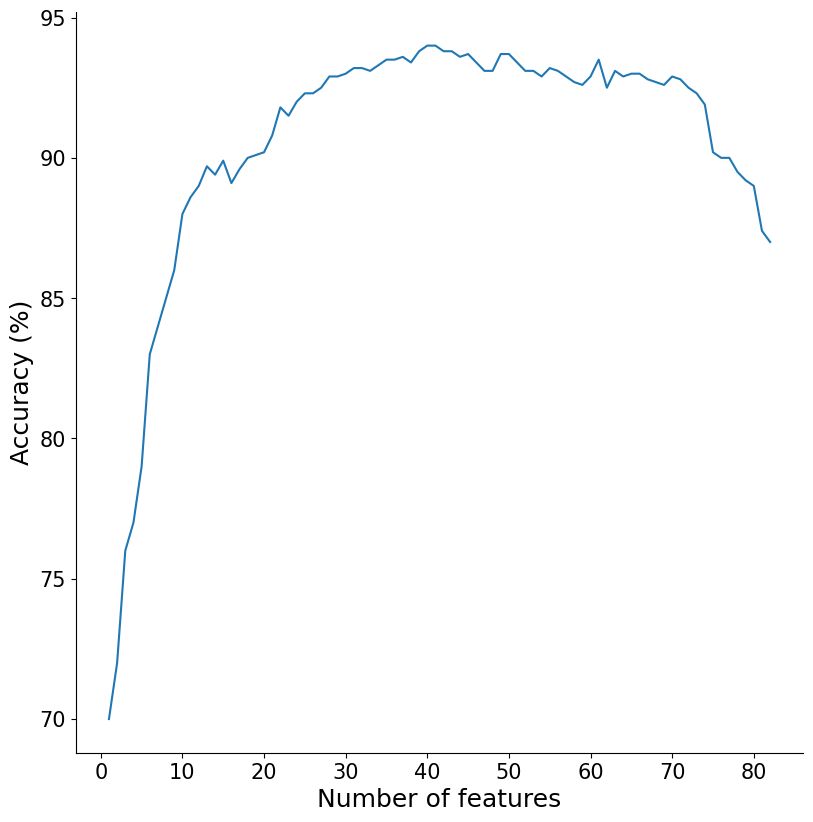}
\caption{ Accuracy obtained in classification of the pulmonary sounds on test set versus the number of features, used in the combined model. Note the decrease in the accuracy, due to over-fitting. }
\label{fig:features_accuracy}
\end{figure}

Finally, Figure~\ref{fig:features_accuracy} presents the accuracy on the test set of the combined model for the multiclass classification, over the number of features selected. The results are computed by using the test set. The most accurate classification results are achieved with 40 features. The accuracy decreases with  higher number of features, as the model begin to overfit the data.

\section{Discussion} \label{sec:discussion}
The success penetration of innovative machine learning methodologies into certain disciplines of medical diagnostics, such as radiology, has increased the interest in achieving an automatic process for analysis of lung sounds.

This a process can be used for second opinion in the diagnosis of pulmonary diseases, it can  be useful for monitoring patients in critical care units and may even substitute for medical experts in mass screening of malaria, as example, in parts of the world where there is severe deficiency in medical expertise and  far from sufficient manpower. 

With these goals in mind, a new approach has been proposed in this work, based on geometric properties of the spectrographic representation of lung sounds.
This model outperformed MFCC-based model in four out of five pathologies for which we had access to sufficient data. Further, as we have shown, the  two approaches can be combined by
concatenating feature vector, or by linear regression between the models output probabilities.
These approaches can even be incorporated into a deep learning framework which has emerged as the most widely used approach in the field of machine learning. Such a hybrid paradigm has recently attracted the attention of the community of AI \cite{rueckert2019model}.

\begin{figure}[h!]
\centering
\includegraphics[width=1\columnwidth]{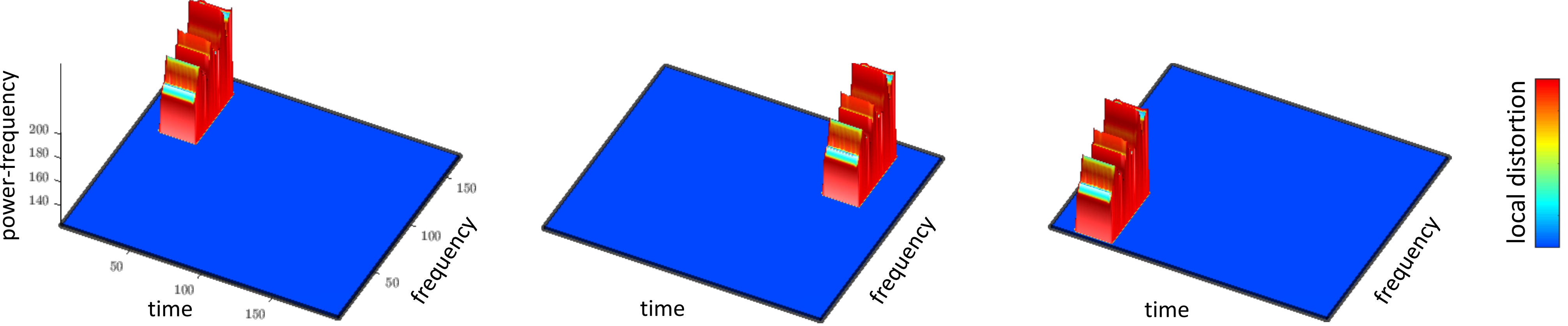}
\caption{
This figure illustrates that, in certain scenarios, distortion measures are invariant under translations of spectrograms in the time-frequency domain.
We show from left to right: a surface representation of the given spectrogram, the same surface shifted along the time and  the frequency axes. 
Colors encode the MIPS distortion \eqref{eq:E_MIPS} induced by the  projection  $f_\text{proj}:(x,y,z)\mapsto (x,y,0)$.
It is easy to show that $f_\text{proj}$  induces equal global distortions over the depicted surfaces, for any local distortion $E(\sigma_1,\sigma_2)$.
Indeed, triangles that belong to the flat regions have zero distortions under $f_\text{proj}$,
whereas triangles of the peaks  produce the same amount of a local distortion because $E(\sigma_1,\sigma_2)$ is invariant under translations.
}
\label{fig:distortion_invariant}
\end{figure}

It should be noted that our model  matches the performance of the existing deep learning methods, while requiring
much fewer data samples for training.  
Furthermore, compared with purely data-driven approaches, our handcrafted features allow a better theoretical understanding of
the sound classification problem. 
In particular, by analyzing properties of simplicial maps we can identify distortion-preserving transformations of audio signals
in the time-frequency domain (see Figure~\ref{fig:distortion_invariant} for the illustration). 

With the available, rather limited volume of dataset, we cannot  predict what will be the optimal subset of features to select in the combined method that  incorporates MFCC and distortion measures. Indeed, the subset of features varies from disease to disease. A larger dataset is needed to assess what is the global optimal subset of features.

Our novel approach to classification, based on distortion measures, highlights an interesting direction for future research would be to consider a higher dimensional distortion measures. In particular, a spectrogram surface can be represented by the tetrahedral volume enclosed by that surface and the plane $z=0$.  In such a case, sound spectrograms can be characterized by 3D distortions induced by mapping tetrahedral meshes into canonical domains. Although the tetrahedral approach has a higher computational cost, it can yield potentially more accurate results because volumetric distortions can detect both the changes made to the boundary surface and the changes made to the interior volume.  

It should be also interesting to combine our model with  various types of shape descriptors, such as the metric and spectral geometrical features, listed in Section \ref{sec:shape_descriptors}.

We  also plan to examine more methods for discretizing the  spectrogram images.  In particular, a curvature based method  \cite{saucan2008sampling} can be used for  a more accurate sampling of spectrogram  images and for building triangle meshes with an optimal number of vertices.

Finally, we stress that our approach to the classification of one-dimensional signals is also applicable to higher-dimensional signals. Distortion measures can be extended in a straightforward manner to $\mathbb{R}^n$ and to piecewise linear manifolds embedded in $\mathbb{R}^n$,  for any $n\geq 2$ \footnote{See \cite{Naitsat2020_preprint}, for $n$-dimensional variants of distortion measures introduced in Section \ref{sec:local_distortions}}. Indeed, if  $\boldsymbol{f}:\mathbb{R}^n\rightarrow  \mathbb{R}^n$ is a simplicial map and $s$ is an $n$-dimensional simplex, then   a local distortion of $\boldsymbol{f}$ over $s$ can be expressed by a function $E(\sigma_1,\sigma_2,\dots,\sigma_n)$, where   $\sigma_i$ denotes the $i^{\text{th}}$ singular value of the Jacobian matrix $d\boldsymbol{f}_s\in \mathbb{R}^{n \times n}$. So that our distortion-based analysis of surfaces has been extended  to $m$-manifolds embedded in $\mathbb{R}^n$ and to their discrete representations, for any $2 \leq m\leq n$. For instance,  consider a simultaneous recording of  different time-varying signals such as pulmonary sounds, heart rate, oxygen saturation, body plethysmography,  etc. Instead of computing spectrograms  for each signal separately, one can represent a $n$-channel data stream by a 2-manifold  embedded in $\mathbb{R}^n$. The obtained manifold can be  discretized  using the sampling method of \cite{saucan2008sampling} and a Delaunay-based algorithm for triangulation. An extension of our approach to manifolds thereby allows a more general analysis of multichannel biomedical data, collected from various devices. We therefore consider other  applications of the proposed distortion-based model in related fields of biomedical signal processing, medical imaging  and voice recognition.

\bibliography{Distortion_measure}
\bibliographystyle{ieeetr}

\ifCLASSOPTIONcaptionsoff
  \newpage
\fi



%\bibliographystyle{IEEEtran}
\clearpage

\section*{Supplements}

\subsection*{Supplementary Note 1}

For Random Forests classifier, the grid focused on:
\begin{itemize}
\item
Number of estimators (100, 110, 120, 150, 200, 250, 300).
\item
Number of features to consider at every split (could be all features, or just the square of overall features).
\item
Maximum number of levels in tree (from 10 to 110, with pace of 10).
\item
Minimum number of samples required to split a node (2,5,10).
\item
Minimum number of samples required at each leaf node (1,2,4).
\item
Enable/Disable bootstrap.
\end{itemize}

For SVM classifier, the grid contained: 
\begin{itemize}
\item
Regularization parameter, C. The strength of the regularization is inversely proportional to C. An exponential
search was used, from 10-9 to 1013.
\item
Radial Basis Function (RBF) Kernel was used.
\item
Degree of the kernel function (relevant only for polynomial kernel): 1, 2, 3, 4, 5, 10.
\item
Value of gamma. An exponential search was used, from 10-9 to 1013.
\item
Whether to use the shrinking heuristic.
\end{itemize}

Regarding the KNN classifier, the random grid has the following features:
\begin{itemize}
\item
Number of neighbors to use for k-neighbors queries: small numbers were tried, like 3, 5, 7, 11 and then numbers
from 41 to 131, with pace of 10.
\item
Weight function used in prediction. Whether uniform, or distance (meaning weight points by the inverse of their
distance).
\item
The algorithm used. It could be ball tree, kd tree, or brute force.
\item
The power parameter for Minkowski metric, p: 1 for manhattan distance, 2 for Euclidean distance.
\end{itemize}

In the case of Logistic Regression model:
\begin{itemize}
\item
The norm used in the penalization. It can be L1, L2, or elasticnet.
\item
Tolerance for stopping criteria. An exponential search was used, from $10^{-9}$ to $10^{13}$.
\item
Inverse of regularization strength, C: an exponential search was used, from $10^{-9}$ to $10^{13}$.
\item
The algorithm to use in the optimization problem. There are liblinear, lbfgs, sag, saga and newton-cg.
\end{itemize}

Adaboost combines a series of weak learners, with the aim of creating an improved classifier. The weak learners vote for the
final prediction label, and a majority voting is performed. The grid search for the Adaboost classifier is constitued of:
\begin{itemize}
\item
Number of estimators: A linear search was used, from 50 to 200 with step of 10.
\item
Learning rate: An exponential search was used, from $10^{-9}$ to 10.
\item
Base estimator: RandomForest, DecisionTree, SVM, LogisticRegression.
\end{itemize}

%%  I'm not sure if we need to define  Delaunay triangulation. Maybe its enough to cite a relevent  book.
\subsection*{Supplementary:  Delaunay triangulation}

A Delaunay  triangulation of a finite set $P\subset \RR^3 $ is defined as a collection $I$ of triangles, such that:
\begin{itemize}
	\item
	$ \mathit{conv} (\Vertex) = \cup_{T\epsilon \Im} T $
	\item
	$ P = \cup_{T\epsilon \Im} V(T) $
	\item
	For every distinct pair $T , U\in\Im $, the intersection $ T \cap U $ is either a vertex, an edge, or empty.
\end{itemize}

For a finite set $ P\subset\RR^3$, the Delaunay triangulation is unique if the points are in a general position,
which means no 5~points are cospherical. The assumption of unique Delaunay triangulation has been made in
this work.

\end{document}